\documentclass[twocolumn]{spie}  

\usepackage{amsmath,amsfonts,amssymb}
\usepackage{graphicx}
\usepackage{epstopdf}
\usepackage{subfigure}
\usepackage[colorlinks=true, allcolors=blue]{hyperref}

\title{Optical demultiplexing of fractal-structured beams in turbulent atmospheric environments}

\author{Xiaojing Weng and Luat T. Vuong}

\affil{Departments of Mechanical and Electrical and Computer Engineering,\\ University of California at Riverside, Riverside, CA, 92521, USA}

\begin{document} 

\maketitle

\begin{abstract}
When information is spatially repeated in self-similar fractal beam patterns, only a portion of the diffracted beam is needed to reconstruct the kernel data. What is unique to a fractal-encoding scheme is that the image demultiplexing process can be, to a first approximation, easily performed optically. In prior work, we experimentally and numerically study fractal-encoded optical beams and their mid- and far-field propagation without added turbulence. Here, we present preliminary simulations of fractal-encoded beams with high turbulence ($C_n^2 \geq 10^{-14}$ m$^{-2/3}$) where we achieve respectable bit error rates of $10^{-3}$. These results are impressive given that: data with low fractal orders is shown, simple threshold-algorithms are used (i.e., no machine learning), and only a third of the beam, off-axis, is needed. More robust channel encoding is associated with increased fractal orders, larger collection areas, and higher kernel singular value decomposition entropy. 
\end{abstract} 



\section{Introduction}
\begin{figure*}[htb!]
\centering
\includegraphics[width=0.8\linewidth]{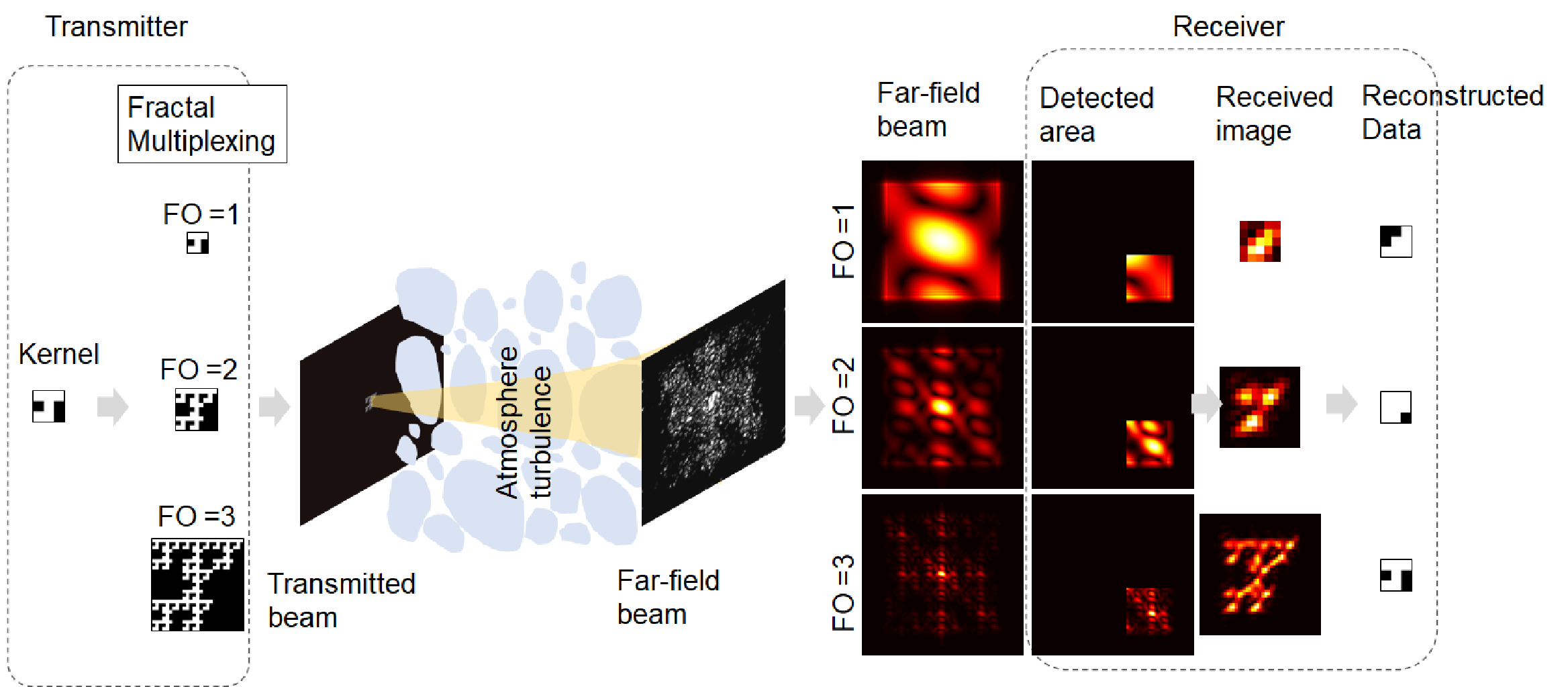}
\caption{Schematic of diffractal space-division-multiplexing (DSDM) with a kernel ``J'' and fractal orders FO = 1, 2, and 3. A partial off-axis portion of the mid-field, diffracted beam is optically demultiplexed with a lens. After a simple threshold algorithm, the FO = 3 data is reconstructed accurately. \textcolor{black}}
\label{fig:diffractal}
\end{figure*}
Although fractals are characterized by high visual complexity, their information content is low: they can easily be generated via simple, recursive algorithms. Mandelbrot elaborated on the fractal self-similar geometry and their mathematical notation in his books \cite{mandelbrot1987fractals}. Following this framework, Berry reported that the diffracted waves from fractal structures exhibit spatiotemporal intensity spiking in their linear propagation dynamics. To emphasize their uniqueness, he referred to the diffracted waves from fractals as `diffractals' \cite{berry1979diffractals}. Fractal geometries and diffractal scattering have attracted widespread attention in many branches of science with applications in engineering such as digital image processing, especially image compression \cite{jacquin1993fractal, zhao2005fractal} and antenna design \cite{puente1998behavior, sharma2017journey, werner2003overview, werner1999fractal, maraghechi2010enhanced}. Such applications exploit a high level of information redundancy, which is organized in strongly-corrugated spatial patterns. An important aspect of diffractals (the diffracted, fractal) is that arbitrary parts of their far-field or Fourier transform  contain sufficient information to recreate the entire sparse, kernel signal. 

In most of the prior experimental work on fractal-shaped beam dynamics, the transmitted beams are digitally encoded or amplitude-modulated. Typically the patterns are Cantor-set, Vicsek fractal patterns, or adapted geometries that are easily calculated via Kronecker-product iterations \cite{Allain1986, Rodrigo2005, AguirreVlez2001}. Still others have theoretically explored the transmission and diffraction of continuous and complex fractal patterns calculated via a cylindrical 2-D Weierstrauss function\cite{Korolenko2021}. Temporal fractal encoding schemes are also proposed \cite{Kavehrad2004}. Towards applications, the facile fabrication of fractal digital masks or grating structures favors intensity-modulated direct detection (IM/DD) schemes \cite{ghassemlooy2019optical}. 

To this end with IM/DD, Cantor-set patterns have been theoretically analyzed in 1-D systems \cite{Verma2013, Verma2012}. In 2-D, the image reconstruction has been experimentally and numerically demonstrated in a 4-F, two-lens system. In this 4-F system, the first lens produces the Fourier-plane field in the focal plane. A tiny aperture in this plane blocks all but a small portion of the field. The aperture-selected fields are reconstructed by a second, Fourier transform lens \cite{moocarme2015robustness}. What is powerful about diffractal propagation is that the central portion of the diffractal is not needed to reconstruct a kernel code and that the original pattern is mostly reconstructed optically. 

In recent work, we numerically study the diffractal space-division multiplexing (DSDM) at non-ideal, mid-field propagation distances \cite{Weng2022}. We observe that even in the mid-field prior to the Fraunhofer diffraction regime, the implementation of a system is robust due to the redundancy built into fractals. With DSDM, the demultiplexing would leverage simple optics. The same simple optics may be used to demultiplex all data channels. Receivers that sample arbitrary beam parts, entirely off-axis are able to reconstruct simple input kernels. There is also a high detection sensitivity: a focusing lens enables capture of low intensity optical signals. Higher-order fractals are more robust, and the capacity to produce higher-order fractals is only limited by number of pixels available at the transmitter. Finally, since the pixel-structured beams exhibit a wide reception cone, DSDM offers a roaming area for a non-coaxial transmitter and receiver. This would aid the tracking and alignment of beam pointing between moving transceivers \cite{Weng2022}. The DSDM transmission of multiple independent or fixed bit streams can also enhance data rates or be used as channel codes under noisy conditions. 

Here, we present preliminary simulations with DSDM in the presence of atmospheric turbulence [Fig. \ref{fig:diffractal}]. Some of these results are posted in a preprint \cite{WengArxiv}. With turbulence ($C_n^2 \geq 10^{-14}$ m$^{-2/3}$), we achieve respectable bit error rates of $10^{-3}$. These results are impressive given that: low fractal orders are shown, simple threshold-algorithms are used (i.e., no machine learning), and only a third of the beam, off-axis, is needed. As previously demonstrated, beams encoded with higher fractal orders achieve higher reconstruction accuracy. We also report that the channel codes with higher robustness appear to also have higher singular value decomposition entropy (a measure of image complexity \cite{Weng2023}). This last trend points to new measures that may aid the design of structured light patterns for free-space communication and sensing.

\section{Optical Demultiplexing}

\begin{figure}[hb!]
\centering
\includegraphics[width=\linewidth]{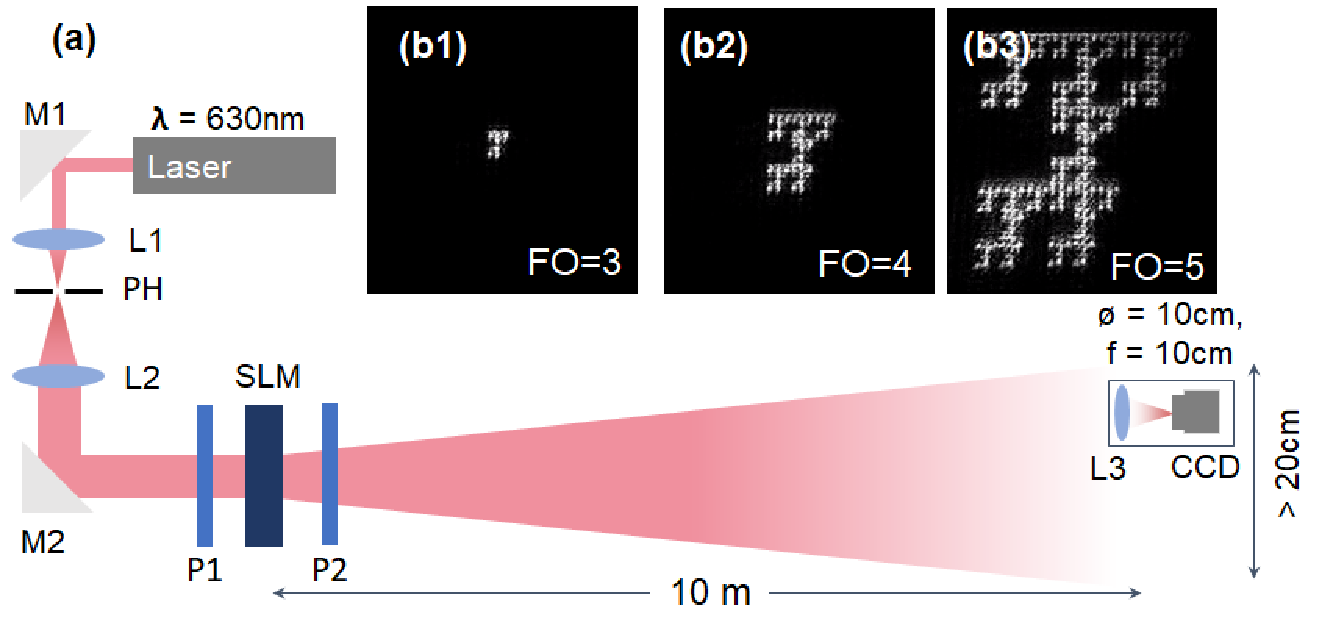}
\caption{(a) Schematic with experimental implementation of DSDM. M1 and M2 are mirrors. L1 and L2 are lenses with pinhole PH to spatially filter. P1 and P2 are orthogonal polarizers \textcolor{black}{ aligned with the spatial light modulator (SLM) with pixel width $a_{pixel} = 36 \mu$m.}  A convex lens L3 with 10-cm focal length focuses and decodes data onto the sensor. (Note: the intensity of the light at the receiver is too low to be measured by the optical camera without the lens L3.) Detector images (b1-b3) show excellent reconstruction of the transmitted beams for fractal orders FO = 3, 4, and 5, in spite of their off-axis locations.}
\label{fig:Experimental}
\end{figure}

In Fig. \ref{fig:Experimental}(a), we show a short-distance experimental setup with a transmissive spatial light modulator (SLM, Holoeye LC2012). The SLM pixels are 36x36 $\mu$m in area. The SLM amplitude-modulates with on/off keying (OOK), where the ‘1’s and ‘0’s are the the presence and absence of light, respectively. For fractal orders (FO) of 3, 4, and 5, the beams transmitted by the SLM are 1, 3, and 9 mm in length, corresponding to 27x27, 81x81, and 243x243 pixel areas. At 10-m distance from the SLM, a fraction of the laser beam is captured off-axis by 10-cm diameter, 10-cm focal-length convex lens. A camera placed in the focal plane of this lens and captures the vignetted Fourier image. As the light is collected further off-axis, we tilt the lens, which produces elongated sensor patterns. The beam diverges over 10-times in width and the light intensity at the receiver is too low to be detected without the focusing lens at the camera sensor. The optically demultiplexed data with FO = 3, 4, and 5 is shown in Figs. \ref{fig:Experimental}(b1-b3), which easily reproduces the transmitted ``J'' kernel. Although the sensor images are elongated, the ``J''-code still visually identifiable. From this image, a simple threshold algorithm would reconstruct the ``J''- kernel code. 

\begin{figure}[b]
\centering
\includegraphics[width=\linewidth]{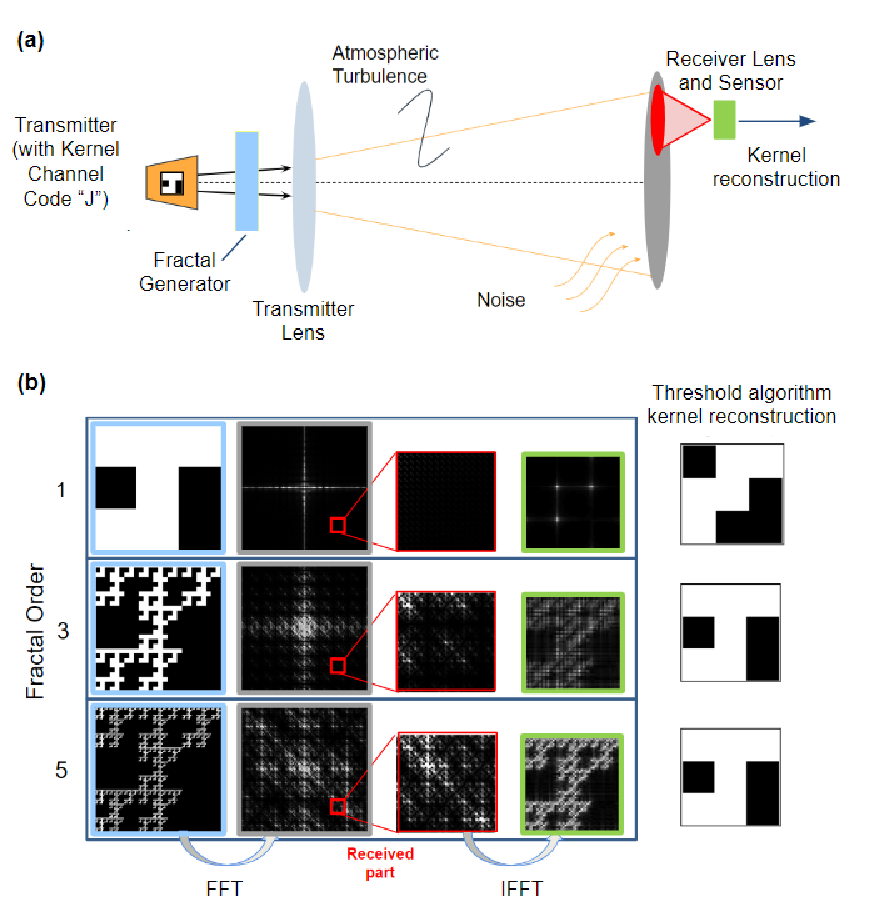}    
\caption{(a) Schematic showing a transmitter with ``J'' kernel code pattern and fractal generator (blue), where a portion of the diffracted field (gray) is sampled and focused with a lens (red) and captured electronically on a sensor (green). (b) The fractal generated output (blue), ideal far-field or Fourier-transform patterns (gray), sampled portion intensity of the far-field (red) and the ideal, reconstructed sensor pattern (green). } 
\label{fig:Schematic3}
\end{figure}

\begin{figure}[b!]
\centering
\includegraphics[width=\linewidth]{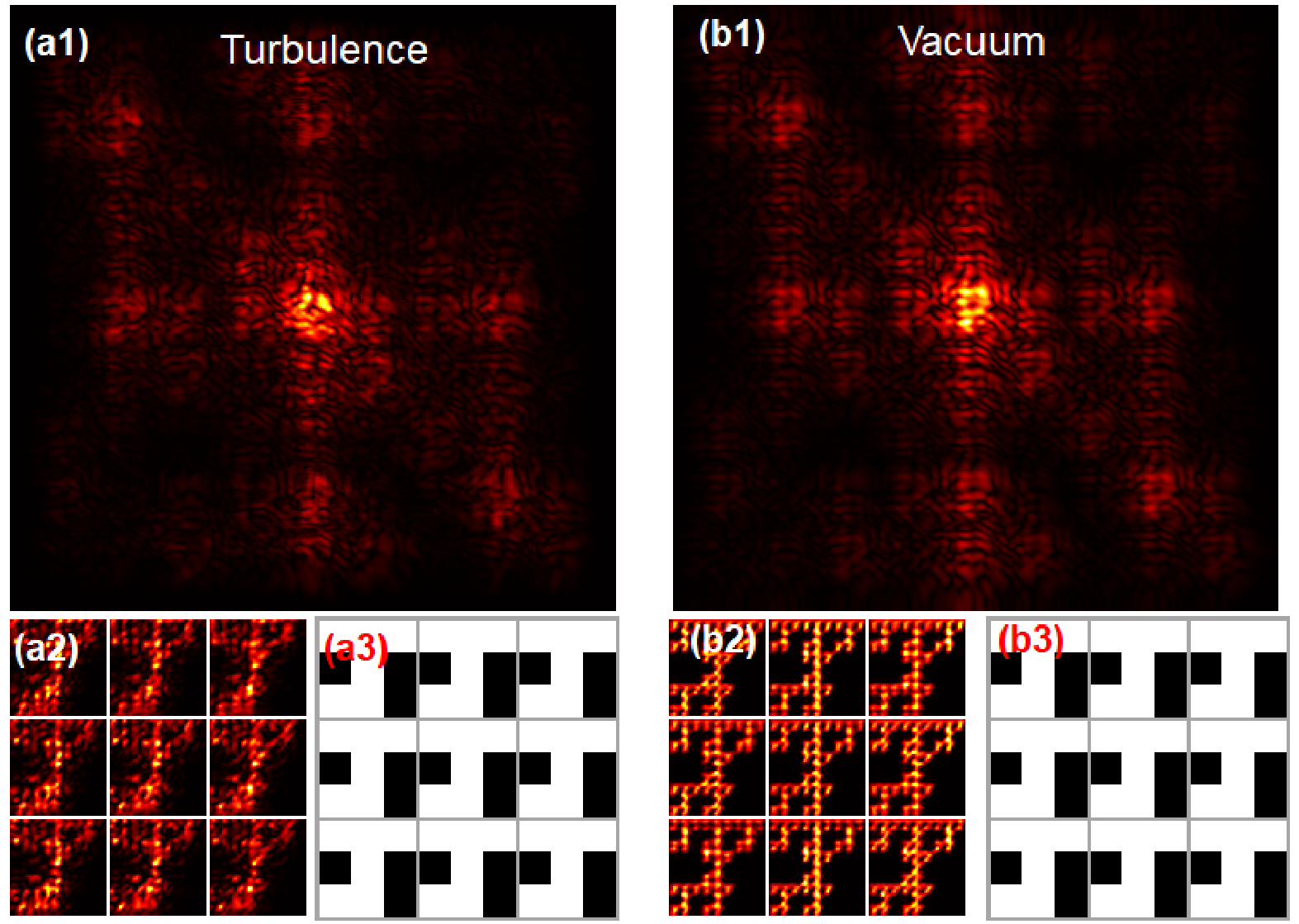}
\caption{Simulated diffraction patterns over an area of 2x2 m at a 2.5-km distance from a transmitted kernel ``J'' with FO = 4 (a1) with turbulence ($C_n^2 = 10^{-13}$ m$^{-2/3}$, inner and outer scales of $l_i=0.01$ m and $l_o=10$ m) and (b1) without turbulence (i.e., in vacuum). (a2, b2) Corresponding sensor patterns for subdivided 3x3, 70-cm-length areas. (a3, b3) Corresponding reconstructed signals from the sensor patterns from a simple threshold algorithm.}
\label{fig:demultiplexing}
\end{figure}



Figure \ref{fig:Schematic3}(a) shows a schematic of how DSDM functions. A transmitter channel operates with the kernel code. This code is iterated in an transmitted fractal pattern. In spite of the noise and turbulence between the transmitter and receiver plane, only a portion of the field is optically demultiplexed and captured.  Figure \ref{fig:Schematic3}(b) shows an ideal far-field simulation where the far-field is calculated with a padded, fast Fourier Transform. (As mentioned earlier, the ideal far-field or Fourier-transform patterns can be demonstrated in 4-F two-lens experiments \cite{moocarme2015robustness}). The encoding that occurs via diffraction is important: more accurate kernel reconstruction occurs when receivers lie in the far-field of the transmitter. (This said, it is also possible to reconstruct the kernel if the beam has only propagated to the mid-field \cite{Weng2022}.)

The kernel reconstruction is performed electronically where the OOK threshold depends on the received image power. This accuracy depends on the location of the receiver in the far field. To reconstruct a 3x3 kernel, the received image is separated into 9 sub-blocks, and each sub-block represents one bit. The mean and variance of each sub-block and background are thresholded to either ‘1’ or ‘0’. 


\section{Numerical Simulations}

We simulate the diffraction of fractals using an angular spectrum method with random phase screens \cite{khare2020orbital}. The atmospheric turbulence is simulated with split-step simulation methods previously established with the Kolmogorov-von Karman turbulence distribution \cite{schmidt2010numerical}. The primary parameter associated with the atmospheric turbulence is the index of refraction structure $C_{n}^{2}$, which is varied to simulate different turbulence strengths. \textcolor{black}{The Rytov variance, a fundamental scaling parameter that depicts the strength of the wave fluctuations, is defined by $\sigma _{I}^{2}=1.23C_{n}^{2}k^{7/6}L^{11/6}$, where $k=2\pi/\lambda$ is the optical wavenumber, $\lambda$ is the wavelength \cite{khare2020orbital}, and $L$ is the propagation length between the transmitter and receiver. Another scaling parameter, the optical signal-to-noise ratio (OSNR), is defined as:}
\begin{eqnarray}
{\rm OSNR} &=&10\log_{10} \left( \frac{Signal}{Noise} \right)\\
&=&10\log _{10}\left( \frac{\sum_1^N{\sum_1^N{|u_{AT}|^{2}}}}{\sum_1^N{\sum_1^N{| u_{AT}-u_{vac} | ^2}}}\right)
\label{eq:SNR}
\end{eqnarray}
where $u_{AT}$ and $u_{vac}$ are the complex, electric-field profiles of the diffracted beams with and without atmospheric turbulence. Given the diffractive encoding of our system, there is some ambiguity whether the OSNR ought to be calculated at the receiver or sensor plane. We calculate OSNR at the receiver lens plane; the effective OSNR at the sensor plane (i.e., in the focal plane of the receiver lens) is generally calculated to be higher than that calculated in the receiver-lens plane.  

We illustrate the robustness of the fractal-encoded system qualitatively and quantatively. Figure \ref{fig:demultiplexing}(a) shows the 2 x 2-m mid-field intensity pattern of a FO = 4, ``J'' - coded fractal with 2-mm pixels that has traveled 2.5 km. In this case, we use a value of $C_n^2 = 10^{-13} $ m$^{-2/3}$, which represents strong turbulence. The optically demultiplexed patterns for each of the one-ninth field areas are shown in Fig. \ref{fig:demultiplexing}(a2). The corresponding reconstructed signals are shown in Fig. \ref{fig:demultiplexing}(a3). Figure \ref{fig:demultiplexing} (b1-b3) show comparison field intensity patterns without turbulence or noise. Under turbulent propagation conditions, both the diffraction and sensor images are distorted, however, we can still reconstruct the correct kernel signal. In the simulation shown in Fig. \ref{fig:demultiplexing}(a3, b3), each of the kernel codes are reconstructed accurately. 

To quantitatively analyze the reliability of the received image with the thresholded algorithm, we define the kernel bit error rate (K-BER). The K-BER is used to evaluate the image reconstruction accuracy of optical demultiplexing. It is different from a typical bit-error-rate calculation because it counts the bit errors of the received kernel compared to a fixed ground-truth transmitted kernel rather than drawing the statistics over all combinations of 1's and 0's. When the kernel is properly reconstructed, the K-BER is zero. If one block of nine is not accurately constructed, the K-BER is 1. The maximum K-BER for a 3x3 kernel is 9. Statistics for K-BER can be calculated over thousands of blocks to draw statistics of the influence of the atmospheric turbulence. 

\begin{figure}[hb!]
    \centering
    \includegraphics[width=\linewidth]{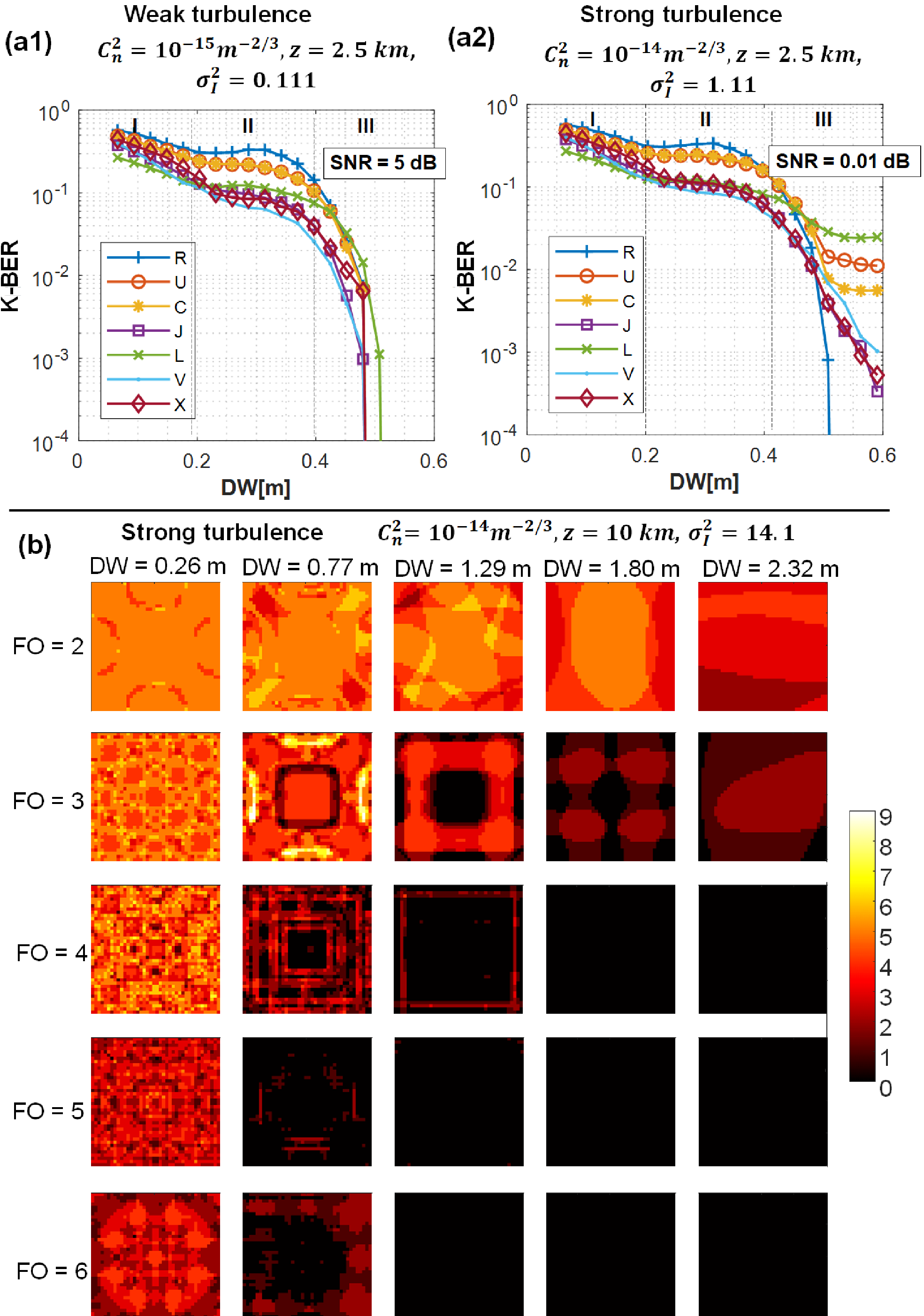}
    \caption{K-BER performance vs. receiver lens width (DW) under (a1) weak turbulence conditions ($C_n^2 = 10^{-15}$ m$^{-2/3}$, $\sigma_I^2=0.11$, and OSNR of 5 dB) and (a2) strong turbulence conditions ($C_n^2 = 10^{-14}$ m$^{-2/3}$, $\sigma_I^2 = 1.11$, and OSNR of 0.01 dB). The fractal order is FO = 4 and the propagation distance is $z = 2.5$ km. (b) The received bit error at different locations within the maximum roaming area with 2.8-m radius with increasing receiver size (DW) by column and and increasing fractal order by row. Here, $z = 10$ km. }
    \label{fig:BERspatially}
\end{figure}


\begin{figure*}[htb!]
    \centering
    \includegraphics[width=1\linewidth]{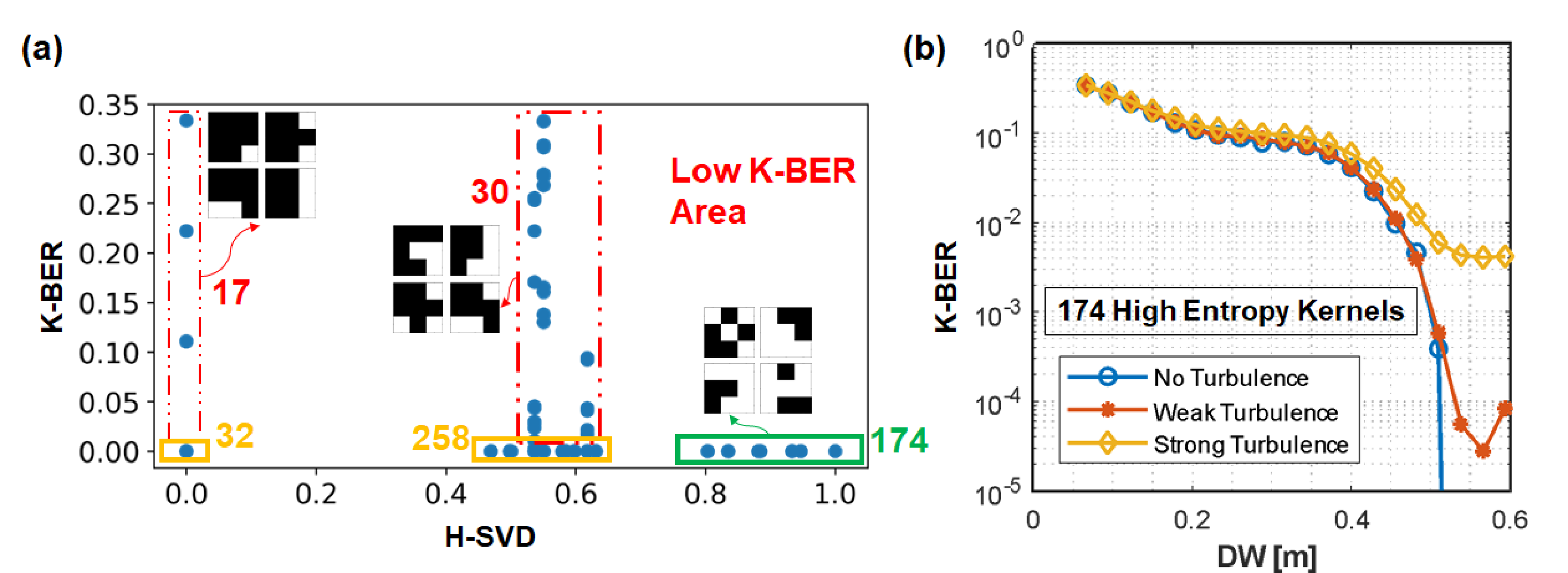}
    \caption{(a) Trend in K-BER performance vs. singular value decomposition entropy (SVD-H) for 511 3x3 kernels. Insets show sample kernels for low, medium, and high values of SVD-H around 0, 0.5, and 0.9. In general, high SVD-H kernels have lower K-BER. (b) K-BER performance vs. receiver lens width (DW) of 174 high-entropy kernels under weak ($C_n^2 = 10^{-15}$ m$^{-2/3}$, $OSNR=3.36$ dB) and strong ($C_n^2 = 10^{-14}$ m$^{-2/3}$, $OSNR=-2.58$ dB) turbulence at $z = 2.5$ km. No machine learning or back-end algorithm is used besides thresholding.}
    \label{fig:Kern174_BER}
\end{figure*}

The K-BER performance under different atmospheric turbulence strength is simulated, calculated and studied as a function of the receiver lens width (DW) [Fig. \ref{fig:BERspatially}(a1,a2)]. In this case, we only consider the error associated with fixed 3x3 kernels that relate the letters, ``R'', ``U'', ``C'', ``J'', ``L'', ``V'', and ``X'' \cite{Weng2022}. The inner and outer lengths associated with the turbulence profile are 0.01 m and 3 m. The propagation distance is $z$=2.5 km, the size of each pixel is 2 mm, and receiver width DW ranges from 5 to 60 cm. The K-BER is calculated from independent simulation runs by averaging 4000 single-aperture receivers (4000 runs) at random locations within the maximum roaming area. Again, in order to use only uncorrelated results, we run a unique simulation with different sets of phase screens for each randomly-detected area. The K-BER performance under weak turbulence is almost the same as that with no turbulence, illustrating that the DSDM system is robust to noise under 5 dB OSNR. With strong turbulence, obvious differences arise in Region III, where receiver width DW $>$ 0.4m. 

Figure \ref{fig:BERspatially}(a1) shows the K-BER under weak turbulence, where $C_n^2 = 10^{-15}$ m$^{-2/3}$, and scintillation index is $\sigma_I^2=0.11$, corresponding to an OSNR of 5 dB. Figure \ref{fig:BERspatially}(a2) shows K-BER under strong turbulence, where $C_n^2 = 10^{-14}$ m$^{-2/3}$, and scintillation index is $\sigma_I^2 = 1.11$, corresponding to a OSNR close to 0.01 dB. Our results are particularly impressive if we consider the low FO and small number of pixels needed to transmit DSDM data. The DSDM channel that transmits a 3x3 kernel data with FO = 4 requires only 81x81 pixels, while an FO = 3 with the same kernel requires only 27x27 pixels. 

\textcolor{black}{The K-BER performance under weak turbulence is almost the same as that with no turbulence. This illustrates that the DSDM system is robust to noise under receiver 5 dB OSNR. With strong turbulence, obvious differences arise in Region III, where receiver width DW $>$ 0.4 m.} 

The reconstruction accuracy generally increases with increasing receiver width DW and also by increasing the FO of the transmitted beam. To illustrate this effect, we plot K-BER values spatially at equally distributed locations within the roaming area at a propagated distance of $z = 10$ km for different FO's [Fig. \ref{fig:BERspatially}(b)]. 

\textcolor{black}{A longer propagation distance is necessary to continue this trend when the fractal order is larger than 4 since a larger FO requires longer distances for diffraction encoding. Unlike previous figures that show the average K-BER from 4000 randomly-positioned receivers, here we show the K-BER from 40x40 single receivers shifted in position and at evenly distributed locations over the roaming area.} In Fig. \ref{fig:BERspatially}(b), different colors represent 0 to 9 received error bit values (there are 9 bits in each kernel). 

By column, receiver widths from 0.26 to 2.32 m are tested. For the same FO, larger receiver widths correspond with fewer error bits, consistent with the declining curves in Fig. \ref{fig:BERspatially}(a1,a2). By row, we show FO from 2 to 6. For the same receiver size, when the FO increases, the bit error decreases. Larger FO's generally improve DSDM robustness to atmospheric turbulence; however, larger FO increases the distance needed for diffraction encoding \cite{Weng2022}. 

The issue of diffraction encoding is highlighted with FO = 4,5,6 at DW = 0.77 m in Fig. \ref{fig:BERspatially}(b). Higher FO's have a smaller K-BER up to FO = 5, but when the FO increases to 6, the corresponding K-BER increases when we expect it to decrease. \textcolor{black}{This increase appears to break the trend where smaller error accompanies higher FO. 
In this case, the distance $z$ = 10 km is significantly less than the minimum diffraction-encoded distance and is not far enough for FO = 6 to reach the ``far-field''. } 

\section{Singular Value Decomposition Entropy}

 \textcolor{black}{While the K-BER describes the kernel reconstruction accuracy, not all kernels are suitable for DSDM. If the size of the data kernel is 3×3, then there are $2^9$ (512) unique combinations of data kernels. For example, the complexity of kernels with only one bit ‘1’, and eight bits ‘0’ always remain the same and independent of FO. An outstanding question is, what is the relationship between reconstruction accuracy and choice of kernel? Here, we use singular value decomposition entropy  (SVD-H) \cite{Weng2023} as a measure of image complexity, defined as: $H_{SVD} = \sum_i{-\sigma _i\log\sigma_i}$, where $\sigma$ are the singular values of the kernel image. We find that more complex patterns measured by kernel SVD-H correlate with lower K-BER. }
 
 Figure \ref{fig:Kern174_BER}(a) shows the K-BER and SVD-H of 511 kernels (the kernel that nine bits are all ‘0’s is excluded). The K-BER values are averaged over 900 random locations when DW = 0.6 m, $z = 2.5$ km, and $C_n^2 = 10^{-15}$ m$^{-2/3}$. When the SVD-H is zero, 32 kernels have K-BER values that are largely zero and 17 kernels have a K-BER larger than 0. When SVD-H is larger than 0.7, there are 174 kernels whose K-BER is largely zero. These 174 high-entropy kernels are selected randomly to study the K-BER under different turbulence conditions [Fig. \ref{fig:Kern174_BER}(b)]. In these cases, the channel code is reconstructed with accuracy of $\log_2 (174) = $7.4 bits per 9-bit frame. Results show that the K-BER performance under weak turbulence is almost the same with that achieved with no turbulence. The K-BER of $10^{-4}$ is achieved when a receiver is about 30\% of maximum roaming area under weak turbulence (3.36 dB OSNR). These results without additional mitigation (machine learning) are used to interpret the signal images. The high SVD-H kernels show outstanding potential for diffraction-encoding and fractal modulation schemes that are robust in adverse environments.


\section{Conclusion}

With DSDM, information is iteratively, self-similarly, and redundantly encoded. With a sufficiently large receiver, sufficiently small pixels and large fractal order, the kernel code is spatially multiplexed over a wide field of view. One advantage of fractal encoding is that the receiver patterns can be demultiplexed optically. This means that any further back-end electronic algorithm that produces the kernel code from the noisy sensor patterns can be simple. 

Since DSDM represents a significantly different diffraction-encoded paradigm for communication and sensing, the calculations for OSNR, BER, and reconstruction accuracy are hard to define and compare with other current communication schemes. Even so, we are able to show impressive results comparing K-BER vs. OSNR with $C_n^2 \geq 10^{-14}$ m$^{-2/3}$ \cite{Willner2021}. Higher FO, smaller pixels, and larger singular value decomposition entropy kernels are more robust given similar adverse conditions.

\section{Backmatter}
LTV gratefully acknowledges funding from DARPA YFA D19AP00036.

\bibliography{reference}
\bibliographystyle{spiebib}

\end{document}